\documentclass[prb,amsmath,showpacs,twocolumn]{revtex4}

\usepackage{amsfonts}
\usepackage{graphicx}

\DeclareMathSymbol{\Lambda}{\mathalpha}{letters}{3}
\DeclareMathSymbol{\Phi}{\mathalpha}{letters}{8}

\DeclareMathOperator{\RE}{Re}
\DeclareMathOperator{\IM}{Im}
\DeclareMathOperator{\e}{e}
\newcommand{\I}{\mathrm{i}}
\newcommand{\diff}{\mathrm{d}}

\newcommand{\vect}[1]{\mathbf{#1}}
\newcommand{\vers}[1]{\hat{\mathbf{#1}}}
\newcommand{\abs}[1]{\lvert #1\rvert}
\newcommand{\tsub}[1]{_{\text{#1}}}
\newcommand{\tsup}[1]{^{\text{#1}}}

\newcommand{\etal}{\textit{et al.}}

\begin{document}

\title{Model of light collimation by photonic crystal surface modes}
\author{Wojciech \'Smigaj}
\email{achu@hoth.amu.edu.pl}
\affiliation{Surface Physics Division, Faculty of Physics, 
  Adam Mickiewicz University,\\
  Umultowska 85, 61-614 Pozna\'n, Poland}

\begin{abstract}
  We propose a quantitative model explaining the mechanism of light
  collimation by leaky surface modes that propagate on a corrugated
  surface around the output of a photonic crystal waveguide.  The
  dispersion relation of these modes is determined for a number of
  surface terminations.  Analytical results obtained on the basis of
  the model are compared to those of rigorous numerical simulations.
  Maximum collimation is shown to occur at frequency values
  corresponding to excitation of surface modes whose wave number
  retains a \emph{nonzero} real part.
\end{abstract}
\pacs{42.70.Qs, 
42.79.Ag, 
78.68.+m
}

\maketitle

\section{Introduction}

One of the problems hindering wider commercial application of photonic
crystals (PCs) is the difficulty in coupling PC waveguides to
conventional dielectric waveguides or optical fibers. A possible
solution consists in tapering the waveguide so as to achieve better
coupling with the fiber; this has been the subject of a number of
publications, e.g., Refs.~\onlinecite{MekisJLT01, HakanssonJLT05}.
\begin{figure}[b]%
  \includegraphics{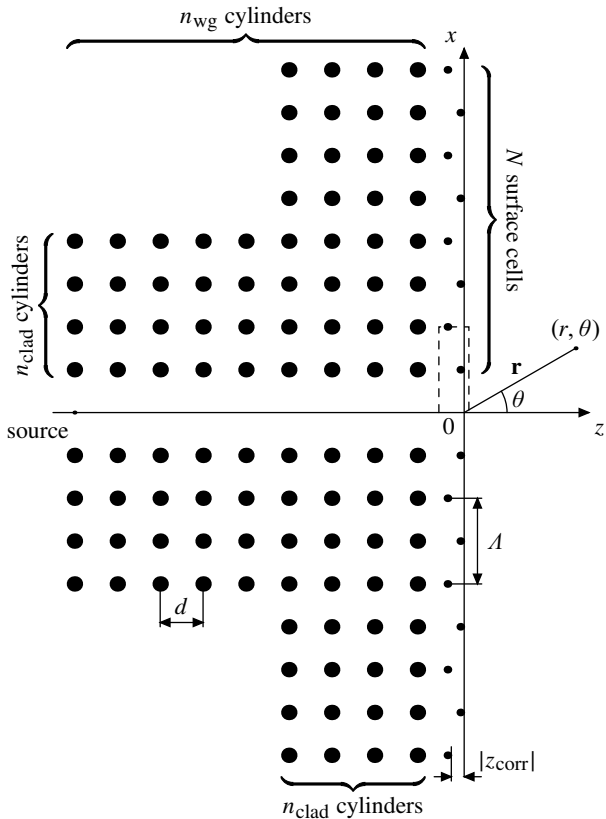}%
  \caption{A waveguide embedded in a PC with corrugated surface.}
  \label{fig:System}
\end{figure}
Recently, Moreno \etal \cite{MorenoPRB04} and Kramper
\etal\cite{KramperPRL04} independently suggested that collimation of
the light emitted by a waveguide (\emph{beaming}) could also occur due
to excitation of surface modes in the proximity of the waveguide exit.
Based on the earlier discovery of a similar effect in metallic
structures supporting surface plasmons,\cite{LezecSci02} the idea has
been expanded in several articles following the original papers.
\cite{MorenoPhNs04, FreiAPL05, MorrisonAPL05, MorrisonApplPhysB05,
  MorrisonSPIE05, BuluOL05}

Moreno \etal\ propose a simple qualitative theory to explain the novel
effect.\cite{MorenoPRB04} The radiation reaching the waveguide outlet
can couple to surface modes; if the surface around the outlet is
\emph{corrugated} (i.e., modulated with period different from that of
the underlying crystal), its eigenstates become `leaky', since energy
is emitted as the radiation scatters at the perturbed surface cells.
Under appropriate conditions, the scattered waves interfere
constructively along the surface normal, thus producing a collimated
beam. According to Ref.~\onlinecite{MorenoPRB04}, this constructive
interference takes place for surface modes of wave vector $k_x = 0$
(shifted to the first Brillouin zone of the surface), with the phase
difference between two successive scatterers equal to an integer
multiple of $2\pi$.

The purpose of this work is to formulate a quantitative model of the
beaming effect in PCs, taking explicitly into account the imaginary
component of the leaky mode wave vector.  The model predictions are
tested against results of numerical simulations. We also show that,
for practically realizable PCs, maximum beaming occurs for surface
modes with $\RE k_x \neq 0$.

\section{Model}
\label{sec:Model}

The system to be considered is depicted in Fig.~\ref{fig:System}.
Excited by a source at the waveguide input (left), a guided mode
propagates towards the crystal surface (right). On reaching the
output, the radiation is partially reflected, partially emitted
directly into free space, and the remainder excites leaky modes
propagating upwards and downwards along the surface corrugated with
period~$\Lambda$. Either side of the waveguide comprises
$N$~surface unit cells. Our aim is to calculate the radiation
intensity $\Phi(\theta)$ defined as
\begin{equation}
  \label{eq:RadiationIntensityDef}
  \Phi(\theta) \equiv \lim_{r\to\infty} r\, S_r(r, \theta),
\end{equation}
where $S_r(r, \theta)$ denotes the radial component of the
time-averaged Poynting vector at the point specified by the polar
coordinates $(r, \theta)$.

Without losing generality, we restrict our attention to
$E$~polarization (with the electric field parallel to the
cylinder~axes). Since all sources of the electromagnetic field are
located in the $z<0$ halfspace, the crystal can be regarded as an
aperture antenna, with radiation pattern proportional to the Fourier
transform of the electric field distribution at the $z=0$
axis:\cite{JullBook81}
\begin{subequations}
  \label{eq:ApertureEquations}
  \begin{equation}
    \label{eq:ApertureField}
    \vect E(r, \theta) \simeq \vers y 
    \e^{\I(k_0 r - \pi/4)} \frac{k_0}{\sqrt{k_0 r}}
    f(k_0 \sin\theta) \cos\theta ,
  \end{equation}
  where 
  \begin{equation}
    \label{eq:FourierTransform}
    f(k_0\sin\theta) \equiv \frac{1}{\sqrt{2\pi}}
    \int_{-\infty}^\infty E_y(x, 0) \e^{-\I k_0 x \sin\theta} \,\diff x,
  \end{equation}
\end{subequations}
$k_0 \equiv \omega/c$ denotes the free-space wave number and $r$~is
assumed to be large compared to the system dimensions. Since the
crystal is symmetric with respect to the $z$~axis, $E_y(x, 0) = E_y(-x, 0)$ and
\begin{equation}
  \label{eq:Symmetry}
  f(k_0\sin\theta) = f^+(k_0\sin\theta) + f^+(-k_0\sin\theta),
\end{equation}
where $f^+(k_0\sin\theta)$ is defined as
\begin{equation}
  \label{eq:f+Def}
  f^+(k_0\sin\theta) \equiv \frac{1}{\sqrt{2\pi}}
  \int_{0}^\infty E_y(x, 0) \e^{-\I k_0 x \sin\theta} \,\diff x,
\end{equation}
The field $E_y(x, 0)$ consists basically of three major components:
the beam stemming directly from the waveguide outlet [$E_y\tsup{wg}(x,
0)$], the leaky mode propagating along the corrugated surface
[$E_y\tsup{surf}(x, 0)$], and the residual radiation extending past
the crystal boundaries [$E_y\tsup{res}(x, 0)$]. In this section, we
focus on the surface mode contribution, assuming $E_y(x, 0) =
E_y\tsup{surf}(x, 0)$; the effect of the other components is discussed
in Section~\ref{sec:Results}.

Neglecting the fringe effects at the surface boundaries, we can apply
the Bloch theorem to the field related to the leaky mode. This yields
\begin{widetext}
\begin{equation}
  \label{eq:SurfaceField}
  E_y\tsup{surf}(x, 0) = 
  \begin{cases}
    u(\abs{x}-\frac{d}{2}) \e^{\I k_x(\abs{x} - d/2)} & 
    \text{if $0 < \abs{x} - \frac{d}{2} < N\Lambda$,}\\
    0 & \text{otherwise.}
  \end{cases}
\end{equation}
Function $u(x)$ is $\Lambda$-periodic and $k_x \equiv k_x' + \I k_x''$
($k_x'' > 0$) denotes the leaky mode wave number. We assume the
waveguide output has effective width~$d$, which is the lattice
constant of the underlying PC, and do not consider this area to belong
to the crystal surface. With Eq.~\eqref{eq:SurfaceField} substituted
into Eq.~\eqref{eq:f+Def}, we get through integration:
\begin{equation}
  \label{eq:gCalculation}
  \begin{split}
    f^+(k_0\sin\theta) &= 
    \Bigl[\sum_{n=0}^{N-1} \e^{\I(k_x-k_0\sin\theta)n\Lambda}\Bigr] 
    \e^{-\I k_0 d \sin\theta/2} F(k_0\sin\theta)\\
    &= \frac{1 - \e^{\I(k_x - k_0 \sin\theta)N\Lambda}}
    {1 - \e^{\I(k_x - k_0 \sin\theta)\Lambda}} 
    \e^{-\I k_0 d \sin\theta/2} F(k_0\sin\theta),
  \end{split}
\end{equation}
\end{widetext}
where the \emph{structure factor} $F(k_0\sin\theta)$ is defined as
\begin{equation}
  \label{eq:StructureFactor}
  F(k_0\sin\theta) \equiv \frac{1}{\sqrt{2\pi}}
  \int_0^\Lambda u(x) \e^{\I(k_x-k_0\sin\theta)x} \diff x.
\end{equation}
Being a periodic function, $u(x)$ can be Fourier-expanded:
\begin{equation}
  \label{eq:uFourier}
  u(x) = \sum_{n=-\infty}^\infty u_n \e^{2\pi\I n x/\Lambda},
\end{equation}
resulting in the following form of the formula for $F(k_0\sin\theta)$:
\begin{equation}
  \label{eq:FExplicit}
  \begin{split}
    F(k_0\sin\theta) &= \frac{1}{\sqrt{2\pi}} \sum_n u_n 
    \int_0^\Lambda \e^{\I(k_x + 2\pi n / \Lambda - k_0\sin\theta)x}
    \diff x \\
    &= \frac{1}{\sqrt{2\pi}} 
    \sum_n u_n \frac{\e^{\I(k_x - k_0\sin\theta)\Lambda} - 1}
    {\I(k_x + 2\pi n/\Lambda - k_0\sin\theta)}.
  \end{split}
\end{equation}
In the first approximation, which is often used in analytical
treatment of leaky-wave antennas,\cite{SchweringIEEE83} only the term
with denominator of the smallest magnitude (the zeroth term in the
case of near-zero $k_x$) needs to be kept in the above sum.
Substitution of this approximate structure factor into
Eq.~\eqref{eq:gCalculation} leads to
\begin{equation}
  \label{eq:Approximatef+}
  f^+(k_0\sin\theta) =
  \frac{\I u_0}{\sqrt{2\pi}} \frac{1- \e^{\I(k_x - k_0 \sin\theta)N\Lambda}}
  {k_x - k_0\sin\theta} \e^{-\I k_0 d \sin\theta/2}.
\end{equation}
From the Maxwell equations it can be shown that in vacuum
\begin{equation}
\label{eq:poynting}
S_r(r, \theta) = \frac{1}{2Z_0} \abs{E(r, \theta)}^2 
\quad \text{with} \quad 
Z_0 \equiv \sqrt{\frac{\mu_0}{\epsilon_0}}.
\end{equation}
Therefore, using Eqs.\
\eqref{eq:ApertureField}, \eqref{eq:Symmetry},
\eqref{eq:Approximatef+}--\eqref{eq:poynting}, and the
definition~\eqref{eq:RadiationIntensityDef}, we arrive at
\begin{equation}
  \label{eq:RadiationIntensity}
  \begin{split}
    \Phi(\theta) &= \frac{k_0}{4\pi Z_0} \abs{u_0}^2 \cos^2\theta\\
    &\quad\times
    \biggl\lvert
    \frac{1- \e^{\I(k_x - k_0 \sin\theta)N\Lambda}}
    {k_x - k_0\sin\theta} +
    \frac{1- \e^{\I(k_x + k_0 \sin\theta)N\Lambda}}
    {k_x + k_0\sin\theta}
    \biggr\rvert^2.
  \end{split}
\end{equation}

The unknown coefficient~$u_0$ can be determined from the principle of
energy conservation. Consider a semi-infinite section of the crystal
surface along which the leaky mode in question propagates. By a
procedure analogous to that followed above, we obtain the radiation
intensity generated by the leaky wave in this configuration:
\begin{equation}
  \label{eq:ApertureFieldSemiinf}
  \Phi(\theta) = 
  \frac{k_0}{4\pi Z_0} \abs{u_0}^2 
  \frac{\cos^2\theta}{\abs{k_x - k_0\sin\theta}^2}.
\end{equation}
Integrated over the interval $[-\frac\pi2, \frac\pi2]$, this yields
the total power radiated into free space, which must be equal to the
power $P_0$ exciting the leaky mode. Therefore,
\begin{equation}
  \label{eq:normalized-power}
    P_0 = 
    \frac{k_0}{4\pi Z_0} \abs{u_0}^2 
    \int_{-\pi/2}^{\pi/2} \frac{\cos^2\theta}{\abs{k_x - k_0\sin\theta}^2}
    \,\diff\theta.
\end{equation}
The integral in the above equation can be evaluated analytically.
Thus,
\begin{equation}
  \label{eq:u0Norm}
  \abs{u_0}^2 = \frac{4\pi Z_0 k_0 P_0}{ \,J(k_x/k_0)},
\end{equation}
where
\begin{widetext}
  \begin{equation}
    \label{eq:LargeIntegral}
    \begin{split}
      J(\kappa) \equiv
        \pi + \frac{1}{\I \IM \kappa} 
        \biggl\{&\sqrt{\kappa^2 - 1} 
          \biggl[\arctan\frac{1-\kappa}{\sqrt{\kappa^2 - 1}} - 
           \arctan\frac{1+\kappa}{\sqrt{\kappa^2 - 1}}\biggr]\\
         &{}-\sqrt{(\kappa^*)^2 - 1} 
          \biggl[\arctan\frac{1-\kappa^*}{\sqrt{(\kappa^*)^2 - 1}} - 
           \arctan\frac{1+\kappa^*}{\sqrt{(\kappa^*)^2 - 1}}\biggr]\biggr\}.
    \end{split}
  \end{equation}
\end{widetext}
By substituting Eq.~\eqref{eq:u0Norm} into
Eq.~\eqref{eq:RadiationIntensity}, we obtain the final formula for the
radiation intensity in the system shown in Fig.~\ref{fig:System},
normalized to the accepted power~$P_0$, and expressed solely in terms
of the leaky surface mode parameters:
\begin{equation}
  \label{eq:FinalRadiationDensity}
  \begin{split}
    \frac{\Phi(\theta)}{P_0} &= \frac{k_0^2 \cos^2\theta}{J(k_x/k_0)}\\
    &\quad\times
    \biggl\lvert
    \frac{1- \e^{\I(k_x - k_0 \sin\theta)N\Lambda}}{k_x - k_0\sin\theta}
    +\frac{1- \e^{\I(k_x + k_0 \sin\theta)N\Lambda}}{k_x + k_0\sin\theta}
    \biggr\rvert^2.
  \end{split}
\end{equation}

\section{Numerical determination of~surface modes}
\label{sec:Numerical}

To apply the above-discussed model to a specific photonic surface,
e.g., for the determination of the frequency most conducive to
beaming, it is necessary to calculate the dispersion relation of the
modes supported by the surface. In this section we shall briefly
outline the method we employed for this purpose.

In our approach, we consider a semi-infinite PC with possible surface
reconstruction. The whole system is divided into three parts: the
homogeneous region, the surface, and the underlying semi-infinite, but
otherwise ideal, photonic crystal. The electromagnetic field in the
homogeneous material is represented as a Rayleigh expansion, i.e., a
linear combination of discrete plane waves, whereas in the
semi-infinite crystal the field is expanded into the eigenmodes of the
corresponding infinite structure (a procedure suggested by Istrate
\etal \cite{IstratePRB05}).  Since we are searching for states leaking
energy \emph{away from the surface}, in both regions we only consider
waves that propagate or decay in this direction; the precise rules of
choosing these waves are specified in the Appendix. The complex band
structure necessary to find the field representation in the PC is
calculated by the differential method (see Ref.~\onlinecite{PopovAO00}
for details).

The fields in the homogeneous region and in the PC are linked by the
scattering matrix\cite{PopovAO00} of the surface layer, which
provides the necessary boundary conditions. This leads to a
homogeneous system of linear equations, which must have a non-trivial
solution for surface states to exist. The search for surface modes is
thus reduced to a search for roots of the determinant of a matrix
dependent on $k_x'$, $k_x''$ and~$\omega$.

\section{Results}
\label{sec:Results}

\subsection{Surface mode dispersion} 

\begin{figure}
  \includegraphics{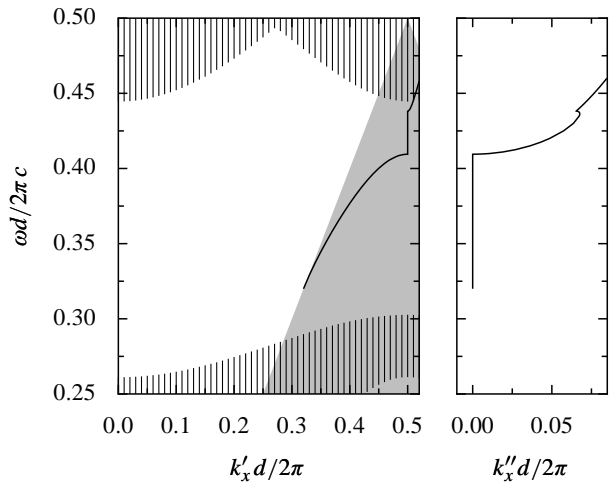}
  \caption{Crystal~$A$ surface mode dispersion curve. The hatched
    areas represent bulk bands, and the grey triangle denotes the
    bound-wave region, in which all spatial harmonics are evanescent
    in vacuum for $k_x'' = 0$.}
  \label{fig:DispersionA}
\end{figure}

\begin{figure}
  \includegraphics{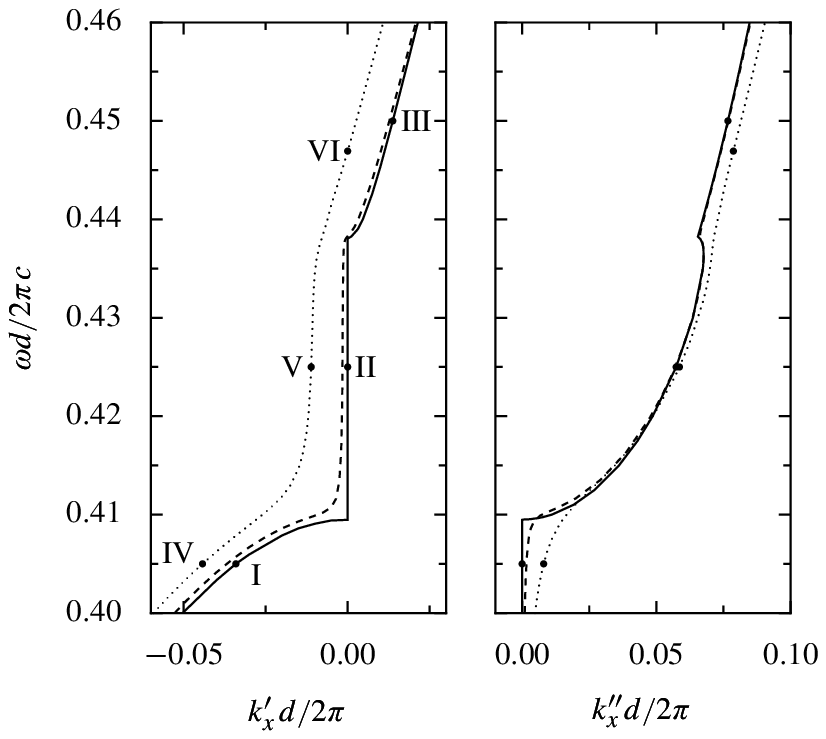}
  \caption{Crystal $B$ and~$C$ surface mode dispersion curves (dashed
    and dotted lines, respectively) in the vicinity of the center of
    the first Brillouin zone. The solid line represents a segment of
    the crystal~$A$ dispersion curve from Fig.~\ref{fig:DispersionA},
    shifted by $\Delta k_x = -\pi/d$ and shown for comparison.}
  \label{fig:DispersionBC}
\end{figure}

In the following we shall focus on crystals of the type shown in
Fig.~\ref{fig:System}, considering a truncated square lattice of
dielectric cylinders of permittivity $\epsilon = 11.56$ embedded in
vacuum, with bulk cylinder radius $0.18d$, surface cylinder radius
$0.09d$, surface corrugation period $\Lambda = 2d$, and three values
of corrugation depth $z_{\text{corr}}$: $0$, $-0.1d$ and $-0.3d$ (the
minus sign indicating that the perturbed cylinders are shifted
\emph{towards} the crystal).  For future reference, we denote these
three crystals with letters $A$, $B$, and~$C$, respectively.
Fig.~\ref{fig:DispersionA} presents the dispersion curve of surface
modes supported by crystal~$A$ (in the extended Brillouin zone
scheme), while in Fig.~\ref{fig:DispersionBC} the dispersion curves
corresponding to crystals $B$ and~$C$ are plotted in the vicinity of
the center of their first Brillouin zone. All these dispersion
relations have been determined by the method described in the previous
section.  Only the modes with $k_x'' \ge 0$ are shown.

\begin{figure}
  \includegraphics{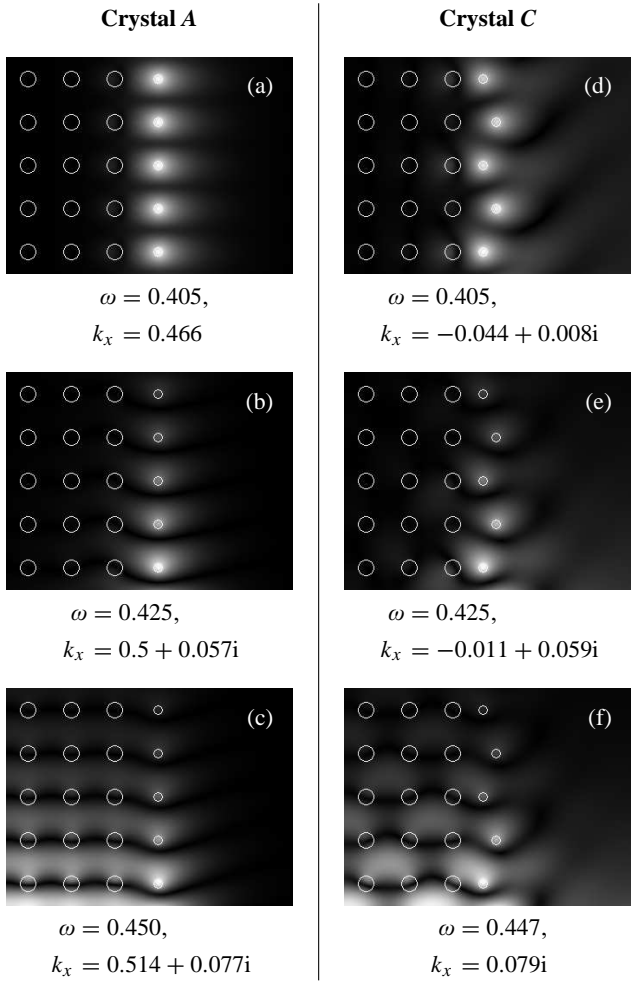}
  \caption{(a)--(f) Maps of electric field magnitude for surface modes labeled
    I--VI in Fig.~\ref{fig:DispersionBC}. The frequency and wave
    number values are expressed in units of $2\pi c/d$ and $2\pi/d$,
    respectively.}
  \label{fig:Maps}
\end{figure}

In the frequency range $\omega < 0.409 \times 2\pi c/d$ crystal~$A$
supports proper (non-leaky, i.e., with $k_x''=0$) surface waves, since
this fragment of the dispersion curve lies in the bounded-wave region
(shaded in Fig.~\ref{fig:DispersionA}), in which all spatial harmonics
are evanescent in vacuum.  Figure~\ref{fig:Maps}(a) shows the field
map of a sample mode from this part of the dispersion curve.  The
range $0.409 < \omega d/2\pi c < 0.438$ corresponds to a stop band,
where the surface mode wave number takes values $\pi/d + \I k_x''$;
see Fig.~\ref{fig:Maps}(b) for the field map of a typical stop-band
state. Characteristic for periodic structures, the occurrence of a
stop band at the Brillouin zone boundary has been observed in
surface-mode dispersion relations of periodic dielectric waveguides
embedded in homogeneous media.\cite{PengIEEE75,DingOE07} In contrast,
the shape of the dispersion curve at $k_x' > \pi/d$ stems directly
from the presence of the underlying PC, i.e., from the periodicity of
the `substrate'.  Although the dispersion curve remains within the
bounded-wave region, the surface mode wave number retains a large
imaginary part, since the wave leaks energy \emph{into the crystal},
as can be observed in Fig.~\ref{fig:Maps}(c), showing the field
magnitude of the mode labeled III in Fig.~\ref{fig:DispersionBC}. This
field structure is analogous to that of leaky modes propagating in
periodic waveguides adjacent to homogeneous media: the wave amplitude
in the PC grows as $z \to -\infty$. As observed in early studies of
leaky waves,\cite{GoldstoneIRE59} this behavior is not unphysical,
since in real systems the field extent is limited by the location of
the source exciting the leaky wave.

Let us proceed to the case of nonzero corrugation (crystals $B$
and~$C$).  Doubling the surface period results in the first Brillouin
zone folding away to the range $[-\pi/2d, \pi/2d]$; consequently, at
frequency values above $0.25 \times 2\pi c/d$ at least one spatial
harmonic is radiative and the originally bound surface states become
leaky. As shown in Fig.~\ref{fig:DispersionBC}, the surface mode
dispersion curve smoothes out, shifting towards negative~$k_x'$ and
larger positive~$k_x''$, as the corrugation depth increases.
Interestingly, at frequency values above $0.438 \times 2\pi c/d$ the
substrate PC supports bulk states characterized by essentially
imaginary~$k_x$ ($k_x \approx \I k_x'' \gtrsim 0.66\I \times 2\pi/d$)
and \emph{purely real}~$k_z$. In consequence, in all three crystals
considered here, the leaky modes from the immediate vicinity of the
$k_x'=0$ line are propagative in the $-z$~direction. Therefore, this
line serves as a boundary between leaky modes exponentially decaying
($k_x' <0$) and growing ($k_x' > 0$) inside the substrate. Figures
\ref{fig:Maps}(d)--(f) show maps of the field corresponding to points
IV--VI on the dispersion curve of crystal~$C$.

\subsection{Light collimation: frequency dependence}

\begin{figure}
  \includegraphics{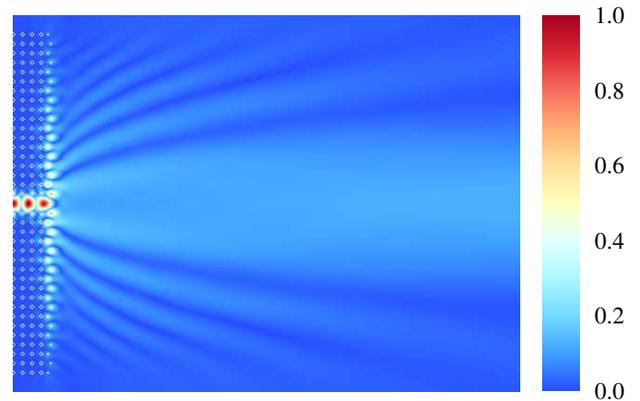}%
  \caption{(Color online) Electric field magnitude for crystal~$C$
    with $N = 9$ corrugated surface cells at frequency
    $\omega = 0.410 \times 2\pi c/d$.}%
  \label{fig:Collimation}%
\end{figure}

\begin{figure}
  \includegraphics{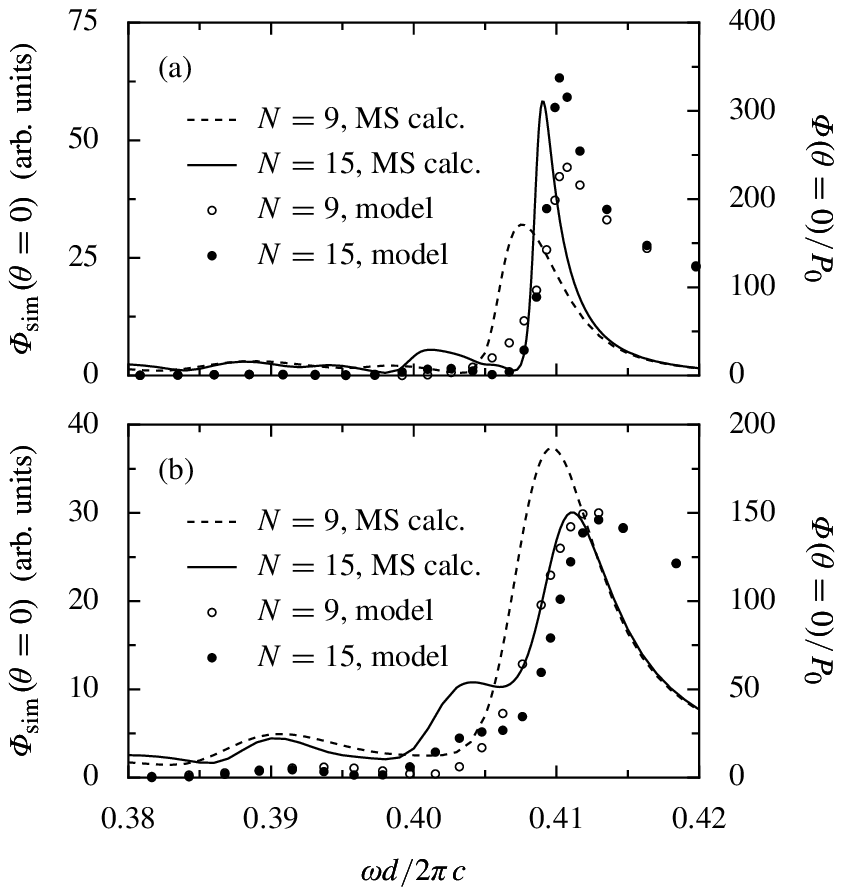}%
  \caption{Frequency dependence of the collimated beam intensity for
    crystal~$B$ (a) and~$C$ (b). The left and right $y$~axes
    refer to the results of MS calculations (lines) and the model
    predictions (circles), respectively.}%
  \label{fig:FreqDep}%
\end{figure}

Let us now proceed to the analysis of beaming itself.
Figure~\ref{fig:Collimation} presents the electric field magnitude
calculated by the multiple-scattering (MS) method (see
Ref.~\onlinecite{FelbacqJOSAA94} for details), with geometry
parameters and frequency value conducive to directional emission. In
Fig.~\ref{fig:FreqDep} the frequency dependence of the radiation
intensity $\Phi_{\text{sim}} (\theta=0)$ calculated by the MS method
is compared to that obtained on the basis of our model, for crystals
$B$ and $C$ with $N = 9$ and $N = 15$ corrugations. In these MS
simulations we consider the system depicted in Fig.~\ref{fig:System},
with the waveguide $n_{\text{wg}} = 12$ cylinders long and cladding
$n_{\text{clad}} = 5$ cylinders wide; a waveguide mode is excited by a
point source near the inlet.  The results depicted in
Fig.~\ref{fig:FreqDep} clearly show that our model reproduces the
basic feature of the effect in question, i.e., the existence of a
distinct transmission maximum at a well-defined frequency value. The
relative height of the $\Phi\tsub{sim}(\theta = 0)$ curve peaks for
different crystals is rendered reasonably well too.  There are some
visible differencies between the model predictions and the simulation
results, though. Most notably, the absolute maxima in the theoretical
curves are shifted by approximately $0.002 \times 2\pi c/d$ to the
right with respect to those found numerically.  Furthermore, at
frequency values smaller (larger) than those corresponding to the main
peaks, the model generally predicts radiation intensity values several
times smaller (larger) than those calculated numerically. Possible
causes of these discrepancies are analyzed in the following subsection
by scrutinizing the angular dependence of the radiation intensity.

\subsection{Light collimation: angular dependence}

\begin{figure}%
  \includegraphics{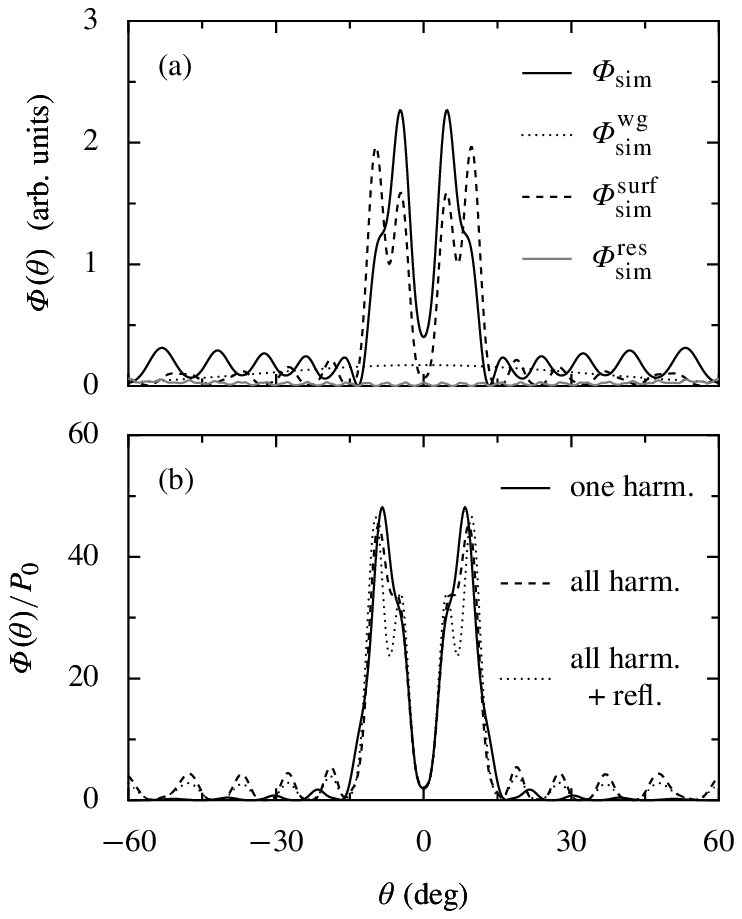}%
  \caption{(a) Angular dependence of the radiation intensity for
    crystal~$C$ with $N = 9$ corrugations at frequency $\omega = 0.400
    \times 2\pi c/d$, calculated by the MS method. Solid black line:
    total radiation intensity; dotted line: contribution of the
    waveguide outlet region, calculated by
    Eq.~\eqref{eq:ApertureEquations} with the integration interval
    restricted to $\abs{x} < d/2$; dashed line: contribution of the
    surface region, with $d/2 < \abs{x} < N\Lambda + d/2$; grey line:
    contribution of the crystal exterior, with $\abs{x} > N\Lambda +
    d/2$. (b)~Angular dependence of the radiation intensity for the
    same crystal and frequency value, calculated on the basis of our
    model with the surface field expansion [Eq.~\eqref{eq:uFourier}]
    truncated to a single harmonic (solid line) and to 17~harmonics
    (dashed and dotted lines). The data plotted with the dotted line
    result from computations taking into account surface wave
    reflections at the crystal boundaries.}
  \label{fig:AngularDep}%
\end{figure}

Figure~\ref{fig:AngularDep}(a) shows the angular dependence of the
radiation intensity $\Phi\tsub{sim}(\theta)$ calculated by the MS
method for crystal~$C$ with $N = 9$ corrugations at frequency $\omega
= 0.400 \times 2\pi c/d$. To help evaluate the relative importance of
the three regions mentioned in Section~\ref{sec:Model}---the waveguide
outlet, the corrugated surface and the crystal exterior---we have also
plotted separately the contributions of sections $\abs{x} < d/2$, $d/2
< \abs{x} < N\Lambda + d/2$ and $\abs{x} > N\Lambda + d/2$, calculated
by the aperture formula~\eqref{eq:ApertureEquations} with the values
of $E_y(x, 0)$ obtained by the MS method. The radiation intensity
produced by each of these regions will be labelled
$\Phi\tsub{sim}\tsup{wg}$, $\Phi\tsub{sim}\tsup{surf}$ and
$\Phi\tsub{sim}\tsup{res}$, respectively.

The graph shows that while surface modes play the most important part
and are responsible for the formation of the major lobes, the
radiation coming directly from the waveguide outlet has an impact as
well. In particular, it makes the main beams shift by several degrees
towards the surface normal. As a result, they begin to overlap at
frequency values below those predicted by leaky-wave considerations
alone. This explains the discrepancy in position of the absolute
maxima between the theoretical and the numerical curves in
Fig.~\ref{fig:FreqDep}, mentioned in the previous subsection.

The radiation pattern in the large-angle ($\theta \gtrsim 15^\circ$)
region is also affected by the field stemming from the outlet: the
destructive interference of this field with that emitted from the
corrugated surface causes an offset of the sidelobe positions.  The
contribution of the residual radiation, however, remains negligible
throughout the angular range covered by the plots in
Fig.~\ref{fig:AngularDep}.

Let us compare the results presented in Fig.~\ref{fig:AngularDep}(a)
with the predictions of our model, plotted in
Fig.~\ref{fig:AngularDep}(b). The solid curve has been
\begin{figure}%
  \includegraphics{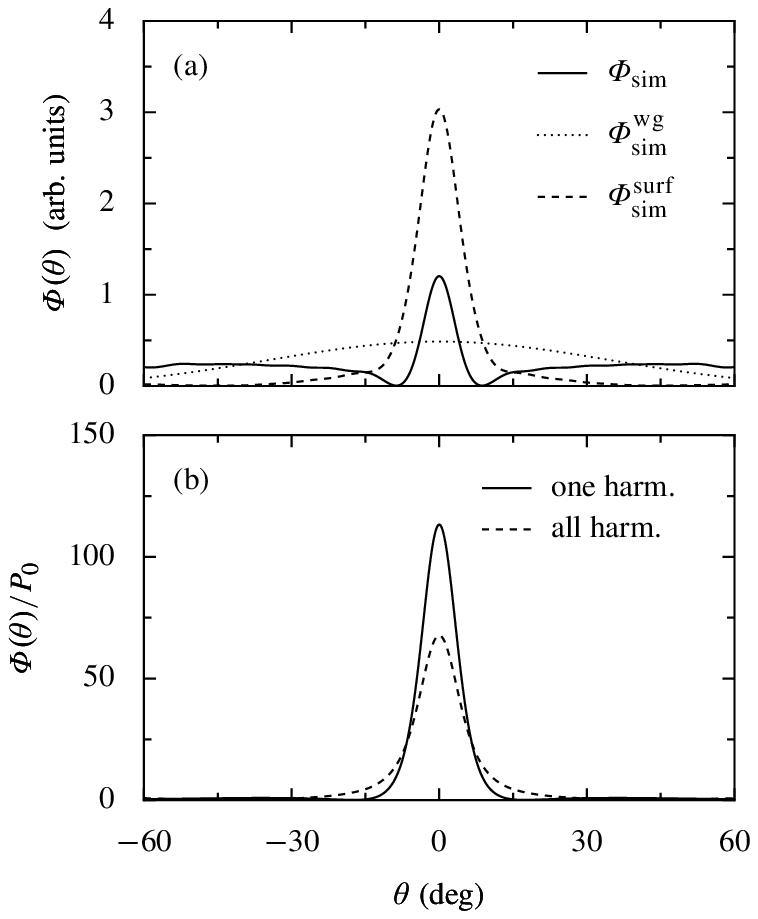}%
  \caption{Same as Fig.~\ref{fig:AngularDep}, but at frequency $\omega
    = 0.420 \times 2\pi c/d$. Since at this frequency virtually no
    power reaches the crystal boundaries,
    $\Phi\tsup{res}\tsub{sim}(\theta)$ and the reflected surface wave
    contribution are not calculated.}
  \label{fig:AngularDep420}%
\end{figure}
calculated with the approximate leaky mode structure factor [i.e.,
with the series in Eq.~\eqref{eq:uFourier} truncated to a single term,
as discussed in Section~\ref{sec:Model}]; the dashed line results from
calculations with the full structure factor (with amplitudes of the
individual harmonics computed numerically by the method outlined in
Section~\ref{sec:Numerical}). In both curves, the main peaks occur at
$\theta_0 \approx 11^\circ$, in good agreement with the
$\Phi\tsub{sim}\tsup{surf}$ curve plotted in
Fig.~\ref{fig:AngularDep}(a). However, the field structure at angles
far from $\theta_0$ is seen to depend strongly on the structure
factor; in general, the radiation intensity obtained with the
one-harmonic approximation is much smaller than that calculated
without this simplification. This may be the reason why at low
frequency values the model-predicted $\Phi(\theta = 0)$ value is very
small compared to that resulting from MS calculations, since for these
frequency values the surface normal lies far from the direction
$\theta = \theta_0$.

\begin{figure*}%
  \includegraphics[scale=0.35]{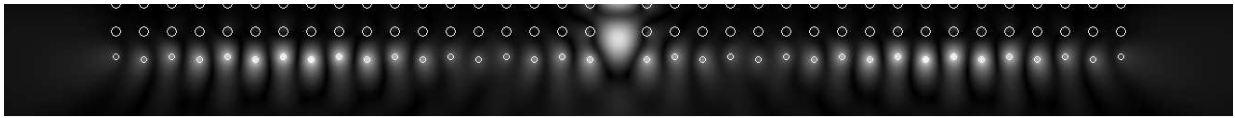}%

  \caption{Electric field magnitude in the surface region of
    crystal~$B$ with $N=9$ corrugations at frequency $\omega = 0.405
    \times 2\pi c/d$. An interference pattern resulting from surface
    wave reflections at the crystal boundaries is clearly visible.}%
  \label{fig:Reflections}%
\end{figure*}

Another difference between the theoretical plots in
Fig.~\ref{fig:AngularDep}(b) and the $\Phi\tsub{sim}\tsup{surf}$ curve
is a distinct dip of the latter at $\theta \approx 6^\circ$. This
turns out to follow from surface wave reflections occurring at the
crystal boundaries, which have been neglected in our model, but are of
some importance for crystals with shallow corrugations (involving
weakly leaky modes) or of small size; see Fig.~\ref{fig:Reflections}.
The dotted curve in Fig.~\ref{fig:AngularDep}(b) shows the radiation
intensity angular dependence after taking these reflections into
account, with the reflection coefficient at the surface termination
calculated by the method outlined in Ref.~\onlinecite{SmigajAX07}.
Evidently, this curve is in excellent agreement with the results of MS
simulations.

Lastly, let us consider the frequency $\omega = 0.420 \times 2\pi
c/d$, which lies to the right of the main peak in the
$\Phi\tsub{sim}(\theta=0)$ curve shown in Fig.~\ref{fig:FreqDep}(b),
and for which the model-predicted radiation intensity at $\theta=0$ is
too large.  Fig.~\ref{fig:AngularDep420}(a), a counterpart of
Fig.~\ref{fig:AngularDep}(a), shows that at this frequency value the
radiation stemming from the waveguide output has magnitude larger than
in the low-frequency case analyzed previously. However, since in the
small-angle region it is out-of-phase with the field produced by the
leaky modes, the total radiation intensity becomes significantly
smaller than it would be without the direct beam from the waveguide
outlet. In addition, as indicated by Fig.~\ref{fig:AngularDep420},
neglecting higher harmonics in the surface field expansion used in the
model-based calculations leads to some overestimation of the
leaky-mode-induced radiation intensity at $\theta = 0$. Together,
these two factors explain the theory-versus-simulation discrepancy
observed in Fig.~\ref{fig:FreqDep} in the frequency range $\omega >
0.41 \times 2\pi c/d$.

\section{Discussion}

As indicated in Fig.~\ref{fig:FreqDep}, maximum beam collimation
occurs at frequency value of around $0.41 \times 2\pi c/d$, which,
according to Fig.~\ref{fig:DispersionBC}, corresponds to surface modes
with $k_x' \approx -0.01 \times 2\pi/d$ (crystal~$B$) and $k_x'
\approx -0.025 \times 2\pi/d$ (crystal~$C$) \emph{rather than} to
modes from the center of the Brillouin zone ($k_x' = 0$).
Incidentally, the $k_x' = 0$ modes, being delocalized (propagative
inside the crystal), could not be responsible for beaming.  However,
the nonzero real part of the wave vector of the surface modes for
which maximum beaming is observed is easily explained on the basis of
the model discussed in Section~\ref{sec:Model}. It is a consequence of
the competition between the tendency to reduce $\abs{k_x'}$ in order
to obtain better phase-matching of waves emitted from individual unit
cells, and, on the other hand, the negative effect of too large a
$k_x''$ on the effective length of the `grating'. Since in the
negative-$k_x'$ region a decrease in $\abs{k_x'}$ is always
accompanied by an increase in $k_x''$, the most intensive beaming
occurs for moderate (`balanced') values of both parameters.

The model also sheds new light on the fact that in negatively
corrugated crystals the frequency value corresponding to maximum
beaming lies remarkably close to that of the unperturbed surface mode
from the Brillouin zone center. This proves to be a resultant of two
opposing effects. It has been pointed out\cite{MorrisonSPIE05} that
bringing surface cylinders closer to the bulk crystal causes a
blueshift of the surface mode dispersion relation, due to a decrease
of the fraction of electromagnetic energy contained within the
dielectric. However, since beaming occurs at nonzero $k_x'$ values,
the shift starts from an initial frequency value lower than that of
the Brillouin-zone-center mode. As a consequence, the resultant
optimum beaming frequency is close to the original frequency of the
$k_x' = 0$ mode.

The accuracy of the model could be considerably improved by taking
into account the radiation emitted directly from the waveguide outlet.
This, however, would require a detailed investigation of the
interactions between the outlet and the surface cylinders in its
immediate vicinity, as these interactions determine the amount of
power transferred to surface modes and that emitted directly into free
space. An analytical formulation of these effects seems hardly
feasible, though.

\section{Conclusions}

We have presented a quantitative analysis of a model explaining the
effect of surface corrugation on the collimation of radiation leaving
the outlet of a photonic crystal waveguide, on the basis of the
dispersion relation of leaky modes supported by the corrugated
surface. The dispersion relation has been calculated and discussed for
a number of surface terminations. The model has been shown to
reproduce the basic features of the investigated effect, and the
significance of the factors left out of account has been evaluated.
Besides clarifying the conditions necessary for optimum beaming, the
model also explains the relative insensitivity of the maximum
collimation frequency value to the degree of surface modulation.  We
believe our results will contribute to a deeper understanding of the
physical grounds of the beaming effect.

\begin{acknowledgments}
I thank Prof.~Henryk Puszkarski and Dr.~Maciej Kraw\-czyk for numerous
useful discussions and encouragement during my work on this project.
\end{acknowledgments}

\appendix*
\section{The choice of basis states}

In this section we specify the selection rules for the states to be
used as the expansion basis for fields in the homogeneous and
periodically modulated system regions considered in
Section~\ref{sec:Numerical}. Let us begin with the homogeneous region.
Here, the electric field can be written as a Rayleigh expansion, i.e.,
a linear combination of spatial harmonics:
\begin{equation}
  \label{eq:Rayleigh}
  E(x, z) = \sum_n A_n \e^{\I(k_{xn} x + k_{zn} z)},
\end{equation}
where $k_{xn} = k_x + 2\pi n/\Lambda$, and the condition $k_{zn}^2 =
k_0^2 - k_{xn}^2$ holds; $k_x$ and $k_0$ are fixed. For real~$k_x$,
the obvious choice for the sign of $k_{zn}$ is
\begin{subequations}
  \begin{align}
    \label{eq:HomogRealSignPlus}
    k_{zn} > 0 \quad &\text{if} \quad k_{zn}^2 > 0,\\
    \label{eq:HomogRealSignMinus}
    k_{zn}/\I > 0 \quad &\text{if} \quad k_{zn}^2 < 0,
  \end{align}
\end{subequations}
so that the propagating harmonics \eqref{eq:HomogRealSignPlus} carry
energy in the $+z$ direction, and the evanescent ones
\eqref{eq:HomogRealSignMinus} decay with $z \to \infty$. When $k_x$~is
complex, the sign of $k_{zn}$ should be chosen so as to assure
analytical continuity with the $k_x \in \mathbb{R}$ case, i.e.,
\begin{subequations}
  \begin{align}
    \label{eq:HomogComplexSignPlus}
    \RE k_{zn} > 0 \quad &\text{if} \quad \RE k_{zn}^2 > 0,\\
    \label{eq:HomogComplexSignMinus}
    \IM k_{zn} > 0 \quad &\text{if} \quad \RE k_{zn}^2 < 0.
  \end{align}
\end{subequations}
It is easy to show that these rules can be combined into
\begin{equation}
  \label{eq:CommonSelRule}
  \RE k_{zn} + \IM k_{zn} > 0,
\end{equation}
in accordance with Refs.~\onlinecite{NeviereBook80,PengIEEE75}.

In the semi-infinite PC, the field can be expanded in the crystal
eigenstates corresponding to the fixed $k_x$ and $k_0$. For
real~$k_x$, these eigenstates are either purely propagative in the
$z$~direction, with $k_z \in \mathbb{R}$, or purely evanescent, with
$k_z = \I k_z''$ or $k_z = \pm \pi/d + \I k_z''$, where $k_z'' \in
\mathbb{R}$. In the former case, the states to be included in the
expansion are those with negative $z$~component of their energy flux
vector $\boldsymbol{\mathcal{E}}$, and in the latter case, those
decaying when $z$ approaches $-\infty$, i.e., those with $k_z'' <0$.
When $k_x$ is allowed to be complex, the $k_z$ component of a crystal
eigenstate wave vector can take arbitrary complex values too.
However, the eigenmodes can still be classified as `essentially
propagative', with $k_z = k_z' + \I k_z''$ fulfilling the condition
\begin{equation}
  \label{eq:BasPropCond}
  \abs{k_z''} < \abs{k_z'} 
  \quad\text{and}\quad
  \abs{k_z''} < \pi/d - \abs{k_z'},
\end{equation}
and `essentially evanescent' otherwise. States of these two families to
be included in the expansion should then be selected according to the
$\boldsymbol{\mathcal{E}} \cdot \vers{z} < 0$ and $k_z'' < 0$ rules,
respectively.


\begin{thebibliography}{21}
\expandafter\ifx\csname natexlab\endcsname\relax\def\natexlab#1{#1}\fi
\expandafter\ifx\csname bibnamefont\endcsname\relax
  \def\bibnamefont#1{#1}\fi
\expandafter\ifx\csname bibfnamefont\endcsname\relax
  \def\bibfnamefont#1{#1}\fi
\expandafter\ifx\csname citenamefont\endcsname\relax
  \def\citenamefont#1{#1}\fi
\expandafter\ifx\csname url\endcsname\relax
  \def\url#1{\texttt{#1}}\fi
\expandafter\ifx\csname urlprefix\endcsname\relax\def\urlprefix{URL }\fi
\providecommand{\bibinfo}[2]{#2}
\providecommand{\eprint}[2][]{\url{#2}}

\bibitem[{\citenamefont{{Mekis} and {Joannopoulos}}(2001)}]{MekisJLT01}
\bibinfo{author}{\bibfnamefont{A.}~\bibnamefont{{Mekis}}} \bibnamefont{and}
  \bibinfo{author}{\bibfnamefont{J.~D.} \bibnamefont{{Joannopoulos}}},
  \bibinfo{journal}{J. Lightwave Technol.} \textbf{\bibinfo{volume}{19}},
  \bibinfo{pages}{861} (\bibinfo{year}{2001}).

\bibitem[{\citenamefont{{H\aa{}kansson}
  et~al.}(2005)\citenamefont{{H\aa{}kansson}, {Sanchis}, {S\'anches-Dehesa},
  and {Mart\'i}}}]{HakanssonJLT05}
\bibinfo{author}{\bibfnamefont{A.}~\bibnamefont{{H\aa{}kansson}}},
  \bibinfo{author}{\bibfnamefont{P.}~\bibnamefont{{Sanchis}}},
  \bibinfo{author}{\bibfnamefont{J.}~\bibnamefont{{S\'anches-Dehesa}}},
  \bibnamefont{and}
  \bibinfo{author}{\bibfnamefont{J.}~\bibnamefont{{Mart\'i}}},
  \bibinfo{journal}{J. Lightwave Technol.} \textbf{\bibinfo{volume}{23}},
  \bibinfo{pages}{3881} (\bibinfo{year}{2005}).

\bibitem[{\citenamefont{{Moreno}
  et~al.}(2004{\natexlab{a}})\citenamefont{{Moreno}, {Garc{\'{\i}}a-Vidal}, and
  {Mart{\'{\i}}n-Moreno}}}]{MorenoPRB04}
\bibinfo{author}{\bibfnamefont{E.}~\bibnamefont{{Moreno}}},
  \bibinfo{author}{\bibfnamefont{F.~J.} \bibnamefont{{Garc{\'{\i}}a-Vidal}}},
  \bibnamefont{and}
  \bibinfo{author}{\bibfnamefont{L.}~\bibnamefont{{Mart{\'{\i}}n-Moreno}}},
  \bibinfo{journal}{Phys. Rev. B} \textbf{\bibinfo{volume}{69}},
  \bibinfo{eid}{121402(R)} (\bibinfo{year}{2004}{\natexlab{a}}).

\bibitem[{\citenamefont{{Kramper} et~al.}(2004)\citenamefont{{Kramper}, {Agio},
  {Soukoulis}, {Birner}, {M{\"u}ller}, {Wehrspohn}, {G{\"o}sele}, and
  {Sandoghdar}}}]{KramperPRL04}
\bibinfo{author}{\bibfnamefont{P.}~\bibnamefont{{Kramper}}},
  \bibinfo{author}{\bibfnamefont{M.}~\bibnamefont{{Agio}}},
  \bibinfo{author}{\bibfnamefont{C.~M.} \bibnamefont{{Soukoulis}}},
  \bibinfo{author}{\bibfnamefont{A.}~\bibnamefont{{Birner}}},
  \bibinfo{author}{\bibfnamefont{F.}~\bibnamefont{{M{\"u}ller}}},
  \bibinfo{author}{\bibfnamefont{R.~B.} \bibnamefont{{Wehrspohn}}},
  \bibinfo{author}{\bibfnamefont{U.}~\bibnamefont{{G{\"o}sele}}},
  \bibnamefont{and}
  \bibinfo{author}{\bibfnamefont{V.}~\bibnamefont{{Sandoghdar}}},
  \bibinfo{journal}{Phys. Rev. Lett.} \textbf{\bibinfo{volume}{92}},
  \bibinfo{eid}{113903} (\bibinfo{year}{2004}).

\bibitem[{\citenamefont{{Lezec} et~al.}(2002)\citenamefont{{Lezec}, {Degiron},
  {Devaux}, {Linke}, {Martin-Moreno}, {Garcia-Vidal}, and
  {Ebbesen}}}]{LezecSci02}
\bibinfo{author}{\bibfnamefont{H.~J.} \bibnamefont{{Lezec}}},
  \bibinfo{author}{\bibfnamefont{A.}~\bibnamefont{{Degiron}}},
  \bibinfo{author}{\bibfnamefont{E.}~\bibnamefont{{Devaux}}},
  \bibinfo{author}{\bibfnamefont{R.~A.} \bibnamefont{{Linke}}},
  \bibinfo{author}{\bibfnamefont{L.}~\bibnamefont{{Martin-Moreno}}},
  \bibinfo{author}{\bibfnamefont{F.~J.} \bibnamefont{{Garcia-Vidal}}},
  \bibnamefont{and} \bibinfo{author}{\bibfnamefont{T.~W.}
  \bibnamefont{{Ebbesen}}}, \bibinfo{journal}{Science}
  \textbf{\bibinfo{volume}{297}}, \bibinfo{pages}{820} (\bibinfo{year}{2002}).

\bibitem[{\citenamefont{{Moreno}
  et~al.}(2004{\natexlab{b}})\citenamefont{{Moreno}, {Mart{\'{\i}}n-Moreno},
  and {Garc{\'{\i}}a-Vidal}}}]{MorenoPhNs04}
\bibinfo{author}{\bibfnamefont{E.}~\bibnamefont{{Moreno}}},
  \bibinfo{author}{\bibfnamefont{L.}~\bibnamefont{{Mart{\'{\i}}n-Moreno}}},
  \bibnamefont{and} \bibinfo{author}{\bibfnamefont{F.~J.}
  \bibnamefont{{Garc{\'{\i}}a-Vidal}}}, \bibinfo{journal}{Photonics and
  Nanostructures} \textbf{\bibinfo{volume}{2}}, \bibinfo{pages}{97}
  (\bibinfo{year}{2004}{\natexlab{b}}).

\bibitem[{\citenamefont{{Frei} et~al.}(2005)\citenamefont{{Frei}, {Tortorelli},
  and {Johnson}}}]{FreiAPL05}
\bibinfo{author}{\bibfnamefont{W.~R.} \bibnamefont{{Frei}}},
  \bibinfo{author}{\bibfnamefont{D.~A.} \bibnamefont{{Tortorelli}}},
  \bibnamefont{and} \bibinfo{author}{\bibfnamefont{H.~T.}
  \bibnamefont{{Johnson}}}, \bibinfo{journal}{Appl. Phys. Lett.}
  \textbf{\bibinfo{volume}{86}}, \bibinfo{pages}{1114} (\bibinfo{year}{2005}).

\bibitem[{\citenamefont{{Morrison} and
  {Kivshar}}(2005{\natexlab{a}})}]{MorrisonAPL05}
\bibinfo{author}{\bibfnamefont{S.~K.} \bibnamefont{{Morrison}}}
  \bibnamefont{and} \bibinfo{author}{\bibfnamefont{Y.~S.}
  \bibnamefont{{Kivshar}}}, \bibinfo{journal}{Appl. Phys. Lett.}
  \textbf{\bibinfo{volume}{86}}, \bibinfo{pages}{1110}
  (\bibinfo{year}{2005}{\natexlab{a}}).

\bibitem[{\citenamefont{{Morrison} and
  {Kivshar}}(2005{\natexlab{b}})}]{MorrisonApplPhysB05}
\bibinfo{author}{\bibfnamefont{S.~K.} \bibnamefont{{Morrison}}}
  \bibnamefont{and} \bibinfo{author}{\bibfnamefont{Y.~S.}
  \bibnamefont{{Kivshar}}}, \bibinfo{journal}{Appl. Phys. B}
  \textbf{\bibinfo{volume}{81}}, \bibinfo{pages}{343}
  (\bibinfo{year}{2005}{\natexlab{b}}).

\bibitem[{\citenamefont{{Morrison} and
  {Kivshar}}(2005{\natexlab{c}})}]{MorrisonSPIE05}
\bibinfo{author}{\bibfnamefont{S.~K.} \bibnamefont{{Morrison}}}
  \bibnamefont{and} \bibinfo{author}{\bibfnamefont{Y.~S.}
  \bibnamefont{{Kivshar}}}, \bibinfo{journal}{Proc. SPIE}
  \textbf{\bibinfo{volume}{5733}}, \bibinfo{pages}{104}
  (\bibinfo{year}{2005}{\natexlab{c}}).

\bibitem[{\citenamefont{{Bulu} et~al.}(2005)\citenamefont{{Bulu}, {Caglayan},
  and {Ozbay}}}]{BuluOL05}
\bibinfo{author}{\bibfnamefont{I.}~\bibnamefont{{Bulu}}},
  \bibinfo{author}{\bibfnamefont{H.}~\bibnamefont{{Caglayan}}},
  \bibnamefont{and} \bibinfo{author}{\bibfnamefont{E.}~\bibnamefont{{Ozbay}}},
  \bibinfo{journal}{Opt. Lett.} \textbf{\bibinfo{volume}{30}},
  \bibinfo{pages}{3078} (\bibinfo{year}{2005}).

\bibitem[{\citenamefont{Jull}(1981)}]{JullBook81}
\bibinfo{author}{\bibfnamefont{E.~V.} \bibnamefont{Jull}},
  \emph{\bibinfo{title}{Aperture antennas and diffraction theory}}
  (\bibinfo{publisher}{Peter Peregrinus Ltd.}, \bibinfo{address}{New York},
  \bibinfo{year}{1981}), p.~\bibinfo{pages}{13}.

\bibitem[{\citenamefont{{Schwering} and {Peng}}(1983)}]{SchweringIEEE83}
\bibinfo{author}{\bibfnamefont{F.~K.} \bibnamefont{{Schwering}}}
  \bibnamefont{and} \bibinfo{author}{\bibfnamefont{S.~T.}
  \bibnamefont{{Peng}}}, \bibinfo{journal}{IEEE Trans. Microw. Theory Tech.}
  \textbf{\bibinfo{volume}{31}}, \bibinfo{pages}{199} (\bibinfo{year}{1983}).

\bibitem[{\citenamefont{Istrate et~al.}(2005)\citenamefont{Istrate, Green, and
  Sargent}}]{IstratePRB05}
\bibinfo{author}{\bibfnamefont{E.}~\bibnamefont{Istrate}},
  \bibinfo{author}{\bibfnamefont{A.~A.} \bibnamefont{Green}}, \bibnamefont{and}
  \bibinfo{author}{\bibfnamefont{E.~H.} \bibnamefont{Sargent}},
  \bibinfo{journal}{Phys. Rev. B} \textbf{\bibinfo{volume}{71}},
  \bibinfo{eid}{195122} (\bibinfo{year}{2005}).

\bibitem[{\citenamefont{{Popov} and {Bozhkov}}(2000)}]{PopovAO00}
\bibinfo{author}{\bibfnamefont{E.}~\bibnamefont{{Popov}}} \bibnamefont{and}
  \bibinfo{author}{\bibfnamefont{B.}~\bibnamefont{{Bozhkov}}},
  \bibinfo{journal}{Appl. Opt.} \textbf{\bibinfo{volume}{39}},
  \bibinfo{pages}{4926} (\bibinfo{year}{2000}).

\bibitem[{\citenamefont{{Peng} et~al.}(1975)\citenamefont{{Peng}, {Tamir}, and
  {Bertoni}}}]{PengIEEE75}
\bibinfo{author}{\bibfnamefont{S.~T.} \bibnamefont{{Peng}}},
  \bibinfo{author}{\bibfnamefont{T.}~\bibnamefont{{Tamir}}}, \bibnamefont{and}
  \bibinfo{author}{\bibfnamefont{H.~L.} \bibnamefont{{Bertoni}}},
  \bibinfo{journal}{IEEE Trans. Microw. Theory Tech.}
  \textbf{\bibinfo{volume}{23}}, \bibinfo{pages}{123} (\bibinfo{year}{1975});
  \bibinfo{note}{\textbf{24}, 542 (1976)}.

\bibitem[{\citenamefont{{Ding} and {Magnusson}}(2007)}]{DingOE07}
\bibinfo{author}{\bibfnamefont{Y.}~\bibnamefont{{Ding}}} \bibnamefont{and}
  \bibinfo{author}{\bibfnamefont{R.}~\bibnamefont{{Magnusson}}},
  \bibinfo{journal}{Opt. Express} \textbf{\bibinfo{volume}{15}},
  \bibinfo{pages}{680} (\bibinfo{year}{2007}).

\bibitem[{\citenamefont{{Goldstone} and {Oliner}}(1959)}]{GoldstoneIRE59}
\bibinfo{author}{\bibfnamefont{L.~O.} \bibnamefont{{Goldstone}}}
  \bibnamefont{and} \bibinfo{author}{\bibfnamefont{A.~A.}
  \bibnamefont{{Oliner}}}, \bibinfo{journal}{IRE Trans. Antennas Propagat.}
  \textbf{\bibinfo{volume}{7}}, \bibinfo{pages}{307} (\bibinfo{year}{1959}).

\bibitem[{\citenamefont{{Felbacq} et~al.}(1994)\citenamefont{{Felbacq},
  {Tayeb}, and {Maystre}}}]{FelbacqJOSAA94}
\bibinfo{author}{\bibfnamefont{D.}~\bibnamefont{{Felbacq}}},
  \bibinfo{author}{\bibfnamefont{G.}~\bibnamefont{{Tayeb}}}, \bibnamefont{and}
  \bibinfo{author}{\bibfnamefont{D.}~\bibnamefont{{Maystre}}},
  \bibinfo{journal}{J. Opt. Soc. Am. A} \textbf{\bibinfo{volume}{11}},
  \bibinfo{pages}{2526} (\bibinfo{year}{1994}).

\bibitem[{\citenamefont{\'{S}migaj}(2007)}]{SmigajAX07}
\bibinfo{author}{\bibfnamefont{W.}~\bibnamefont{\'{S}migaj}}
  (\bibinfo{year}{2007}), \eprint{arXiv:physics/0703165}.

\bibitem[{\citenamefont{Neviere}(1980)}]{NeviereBook80}
\bibinfo{author}{\bibfnamefont{M.}~\bibnamefont{Neviere}}, in
  \emph{\bibinfo{booktitle}{Electromagnetic theory of gratings}}, edited by
  \bibinfo{editor}{\bibfnamefont{R.}~\bibnamefont{Petit}}
  (\bibinfo{publisher}{Springer Verlag}, \bibinfo{address}{Berlin},
  \bibinfo{year}{1980}), pp. \bibinfo{pages}{123--158}.

\end{thebibliography}

\end{document}